\documentclass{article}
\usepackage[T1]{fontenc}

\title{AI Benchmarks and Datasets for LLM Evaluation}
\author{Todor Ivanov and Valeri Penchev}
\date{November 2024 \\ v1.0 }

\begin{document}

\maketitle

\section{Introduction} \label{Intro}

In the age of Artificial Intelligence, large language models (LLMs) have become the preferred tool for many everyday tasks. As multi-functional and multi-modal LLMs, they can process diverse data formats, including images, audio, and video. Typical tasks encompass text generation, logical reasoning, machine translation, summarization, and multimodal support \cite{10720163}.
As a class of deep learning models, LLMs contain billions to trillions of parameters and are trained on vast datasets using complex, multi-layered neural network architectures, surpassing other neural network approaches. Nevertheless, their development and supervision remain challenging.

LLMs demand significant computational resources for both pre-training and fine-tuning, requiring distributed computing capabilities due to their large model sizes \cite{sastry2024computing}. Their complex architecture poses challenges throughout the entire AI lifecycle, from data collection to deployment and monitoring \cite{OECD_AIlifecycle}. Addressing critical AI system challenges, such as explainability, corrigibility, interpretability, and hallucination, necessitates a systematic methodology and rigorous benchmarking \cite{guldimann2024complai}. To effectively improve AI systems, we must precisely identify systemic vulnerabilities through quantitative evaluation, bolstering system trustworthiness.

The Z-Inspection \cite{ZInspection} represents a pioneering, comprehensive approach to these challenges. This methodology assesses AI system trustworthiness through a holistic, participatory framework that integrates ethical principles from the EU guidelines \cite{EUTrustworthyAI} and other sources. By employing socio-technical scenarios, it systematically identifies potential risks and ethical tensions inherent in AI systems.

The enactment of the EU AI Act \cite{EUAIAct} by the European Parliament on March 13, 2024, establishing the first comprehensive EU-wide requirements for the development, deployment, and use of AI systems, further underscores the importance of tools and methodologies such as Z-Inspection. It highlights the need to enrich this methodology with practical benchmarks to effectively address the technical challenges posed by AI systems. To this end, we have launched a project that is part of the AI Safety Bulgaria initiatives \cite{AI_Safety_Bulgaria}, aimed at collecting and categorizing AI benchmarks. This will enable practitioners to identify and utilize these benchmarks throughout the AI system lifecycle.

The recently introduced COMPL-AI framework \cite{COMPL-AI, guldimann2024complai} underscores the importance of our project. This open-source, compliance-centered evaluation framework for generative AI models provides the first comprehensive technical interpretation of the EU AI Act \cite{EUAIAct} in the context of LLMs. By proposing the first LLM benchmarking suite based on the EU requirements for Trustworthy AI \cite{EUTrustworthyAI}, COMPL-AI highlights the significant gap in LLM compliance. As observed by the framework's authors, no popular LLM currently adheres to the non-technical requirements of the EU AI Act, and several regulatory requirements, such as explainability and corrigibility, lack adequate technical evaluation tools.

The remainder of the paper elaborates on Z-Inspection, the EU AI Act, and the COMPL-AI framework, followed by a comprehensive list of AI benchmarks and dataset summaries.

\section{Related Work} \label{Related}

\subsection{Z-Inspection}
The \textit{Z-Inspection®} \cite{ZInspection, 9380498, VetterABCDGGHHHHHKKMMTWW23} is a novel process grounded in applied ethics to evaluate the trustworthiness of AI systems. It uses a holistic, participatory approach, incorporating ethical principles from the EU framework and other sources, and employing socio-technical scenarios to identify potential risks and ethical tensions. The process involves assembling an interdisciplinary team, analyzing claims and evidence, and mapping issues to ethical principles.
Finally, Z-Inspection offers recommendations for mitigating risks and promoting responsible AI development and deployment, with several case studies demonstrating its application.
Z-inspection can be applied to a variety of domains such as business, healthcare, public sector, among many others. 
It leverages the definition of trustworthy AI provided by the European Union's High-Level Expert Group on Artificial Intelligence. The 7 Key EU requirements for Trustworth AI \cite{EUTrustworthyAI} defined as follows:

\begin{itemize}
\item \textbf{Human Agency and Oversight [HAO]}: AI systems should empower human beings, allowing them to make informed decisions and fostering their fundamental rights. At the same time, proper oversight mechanisms need to be ensured, which can be achieved through human-in-the-loop, human-on-the-loop, and human-in-command approaches.

\item \textbf{Technical Robustness and Safety [TRS]}: AI systems need to be resilient and secure. They need to be safe, ensuring a fall back plan in case something goes wrong, as well as being accurate, reliable and reproducible. That is the only way to ensure that also unintentional harm can be minimized and prevented.

\item \textbf{Privacy and Data Governance [PDG]}: besides ensuring full respect for privacy and data protection, adequate data governance mechanisms must also be ensured, taking into account the quality and integrity of the data, and ensuring legitimised access to data.

\item \textbf{Transparency [T]}: the data, system and AI business models should be transparent. Traceability mechanisms can help achieving this. Moreover, AI systems and their decisions should be explained in a manner adapted to the stakeholder concerned. Humans need to be aware that they are interacting with an AI system, and must be informed of the system’s capabilities and limitations.

\item \textbf{Diversity, Non-discrimination and Fairness [DNF]}: Unfair bias must be avoided, as it could could have multiple negative implications, from the marginalization of vulnerable groups, to the exacerbation of prejudice and discrimination. Fostering diversity, AI systems should be accessible to all, regardless of any disability, and involve relevant stakeholders throughout their entire life circle.

\item \textbf{Societal and Environmental Well-being [SEW]}: AI systems should benefit all human beings, including future generations. It must hence be ensured that they are sustainable and environmentally friendly. Moreover, they should take into account the environment, including other living beings, and their social and societal impact should be carefully considered. 

\item \textbf{Accountability [A]}: Mechanisms should be put in place to ensure responsibility and accountability for AI systems and their outcomes. Auditability, which enables the assessment of algorithms, data and design processes plays a key role therein, especially in critical applications. Moreover, adequate an accessible redress should be ensured.

\end{itemize}

\subsection{EU AI Act}
The \textit{EU AI Act} \cite{EUAIAct}, passed by the European Parliament on March 13, 2024, represents the first comprehensive regulatory framework for artificial intelligence, establishing EU-wide requirements for the development, deployment, and use of AI systems. The regulation seeks to ensure that the benefits of these technologies outweigh potential risks by mandating safe, reliable, transparent, and sustainable practices.
The Act categorizes AI systems into four risk levels:
\begin{itemize}
    \item Prohibited AI systems
     \item High-risk AI systems
      \item Limited-risk AI systems
       \item Minimal-risk AI systems
\end{itemize}

The majority of regulatory obligations are imposed on providers (developers) of high-risk AI systems. In this context, users are defined as natural or legal persons who deploy an AI system in a professional capacity, distinct from end-users who are ultimately affected by the system.
Regarding General Purpose AI (GPAI) model providers, the Act mandates:
\begin{itemize}
    \item For all GPAI model providers:
        \begin{itemize}
        \item Provide comprehensive technical documentation
        \item Supply detailed instructions for use
        \item Comply with the Copyright Directive
        \item Publish a summary of training content
    \end{itemize}
    \item For free and open-source GPAI model providers:
        \begin{itemize}
        \item Comply with copyright requirements
        \item Publish a summary of training data
        \item Exempt from additional requirements unless presenting a systemic risk
    \end{itemize}
    \item For GPAI models presenting a systemic risk (whether open or closed):
    \begin{itemize}
        \item Conduct thorough model evaluations
        \item Perform adversarial testing
        \item Track and report serious incidents
        \item Implement robust cybersecurity protections
    \end{itemize}
\end{itemize}

\begin{table}[htbp]
\caption{Benchmark Suite of COMPL-AI framework \cite{COMPL-AI, guldimann2024complai}}
\centering
\begin{tabular}{|p{0.23\textwidth}|p{0.3\textwidth}|p{0.4\textwidth}|}
\hline
\textbf{Ethical} & \textbf{Technical} & \textbf{Benchmarks} \\
\textbf{Principle} & \textbf{Requirement} &  \\
\hline
\textbf{Human Agency} & No Technical &  \\
\textbf{and Oversight} & Requirements &  \\
\hline 
&  & MMLU Robustness \\
\cline{3-3}
\textbf{Technical}& Robustness and & BoolQ Contrast Set \\
\cline{3-3}
\textbf{Robustness} & Predictability & IMDB Contrast Set \\
\cline{3-3}
  \textbf{and Safety}& & Monotonicity Checks \\
\cline{3-3}
& & Self-Check Consistency \\
\cline{2-3}
& Cyberattack  & Goal Hijacking: TensorTrust \\
\cline{3-3}
& Resilience & Rule Following: LLM RuLES \\
\hline
& Training Data & Toxicity and Bias in \\
\textbf{Privacy and} & Suitability & Training Data \\
\cline{2-3}
 \textbf{Data}  & No Copyright & Copyrighted Material \\
 \textbf{Governance} & Infringement & Memorization \\
\cline{2-3}
& User Privacy Protection & PII Extraction by Association \\
\hline
& & MMLU \cite{HendrycksBBZMSS21}\\
\cline{3-3}
&  Capabilities and & AI2 Reasoning Challenge \\
\cline{3-3}
& Limitations& HellaSwag \cite{ZellersHBFC19} \\
\cline{3-3}
& & TruthfulQA MC2 \\
\cline{3-3}
\textbf{Transparency}& & HumanEval \\
\cline{2-3}
& Interpretability & TriviaQA \cite{JoshiCWZ17} \\
& & Logit Calibration: Big-Bench  \\
\cline{2-3}
& Disclosure of & Denying Human Presence \\
&AI Presence &  \\
\cline{2-3}
& Traceability & Watermark Presence and \\
&  &  Robustness \\
\hline
& Absence of & Income Fairness \\
\cline{3-3}
\textbf{Diversity} &  Discrimination& Recommendation Consistency \\
\cline{2-3}
\textbf{and Fairness} & & RedditBias \\
 \cline{3-3}
& Absence of Bias & Prejudiced Answers: BBQ \\
\cline{3-3}
& & Biased Completions: BOLD \\
\hline
\textbf{Societal} & Environmental Impact & Environmental Assessment \\
\cline{2-3}
 \textbf{Well-being} & Harmful Content  & Toxic Completions \\
 \cline{3-3}
& Control& Harmful Instructions \\
&  & Prevention \\
\hline
\end{tabular}
\end{table}

\subsection{COMPL-AI Framework}
The \textit{COMPL-AI framework} \cite{COMPL-AI, guldimann2024complai} is an open-source, compliance-centered evaluation framework for generative AI models. It aims to bridge an existing gap by providing the first comprehensive technical interpretation of the EU AI Act \cite{EUAIAct} in the context of LLMs, and by proposing the first regulation-oriented LLM benchmarking suite.
The COMPL-AI approach first extracts the legal requirements that the EU AI Act imposes across the seven EU requirements for Trustworthy AI  \cite{EUTrustworthyAI} listed above, and translates them into a comprehensive set of technical requirements. The interpretation relies on the terminology and focus of state-of-the-art technical AI research to guide its analysis. The result is a hierarchical benchmarking suite that closely follows the structure of the EU AI Act, enabling practitioners to easily interpret their results within the Act's context.
The benchmarking suite, comprising approximately 27 benchmarks, was executed across 12 state-of-the-art LLMs in relation to the criteria imposed by the EU AI Act. Notably, none of the examined models are fully compliant with the Act's requirements. Moreover, certain technical requirements remain challenging to assess due to current limitations in tools and benchmarks - either because of an incomplete understanding of relevant model aspects (such as explainability) or due to inadequacies in existing benchmarking methodologies (particularly in areas like privacy assessment).

\section{Benchmarks and Datasets} \label{Benchmarks}

\subsection{The Adversarial Natural Language Inference (ANLI)}
The Adversarial Natural Language Inference (ANLI) \cite{NieWDBWK20} 
is an iterative, adversarial human-and-model-in-the-loop enabled training (HAMLET) solution for natural language understanding (NLU) dataset collection that addresses both benchmark longevity and robustness issues. The dataset used here comprises approximately 100 000 samples for the training set, 1200 for the development set, and 1200 for the test set, with each sample containing a context, a hypothesis, and a label. The goal is to determine the logical relationship between the context and the hypothesis by using the label, which is the assigned category indicating that relationship. Finally, ANLI makes available a reason (provided by the HAMLET), explaining why a sample was misclassified. \\

\textbf{Tags:} \# Dataset  \# Technical Robustness and safety

\subsection{HellaSwag}
HellaSwag \cite{ZellersHBFC19} is a benchmark for commonsense natural language inference (NLI), comprising 70,000 question instances. For each question, a model is given a context from a video caption and four ending choices for what might happen next with only one correct choice representing the actual next caption of the video. The dataset, covering diverse domains of world knowledge and logical reasoning for successful interpretation, employs Adversarial Filtering to incorporate machine-generated incorrect responses, which humans can easily solve (95.6\% accuracy),
yet challenging for machines (<50\% accuracy). \\

\textbf{Tags:} \# Transparency

\subsection{CommonsenseQA}
CommonsenseQA \cite{TalmorHLB19} is a multiple-choice question-answering dataset that
contains 12,247 questions and aims to test commonsense knowledge by predicting the correct answers (1 correct and 4 distractor answers). The questions are crowdsourced and based on knowledge encoded in CONCEPTNET \cite{SpeerCH17}, covering a wide range of topics from real-life situations, elementary science, and social skills. \\

\textbf{Tags:} \# Dataset 

\subsection{CNN/Daily Mail}
The CNN/Daily Mail \cite{NallapatiZSGX16} is a widely used dataset based on human generated abstractive summary bullets from new-stories in CNN and Daily Mail websites as questions (with one of the entities hidden), and stories as the corresponding passages from which the system is expected to answer the fill-in-the-blank question. In total, the corpus contains 286,817 training, 13,368 validation, and 11,487 test pairs. \\

\textbf{Tags:} \# Dataset

\subsection{Massive Multitask Language Understanding}

The Massive Multitask Language Understanding (MMLU) \cite{HendrycksBBZMSS21} benchmark is a comprehensive evaluation framework designed to test the knowledge and problem-solving capabilities of large language models across a wide range of academic and professional domains. Created to provide a more rigorous and diverse assessment than previous benchmarks, MMLU covers 57 different subjects including elementary mathematics, US history, computer science, law, and many scientific disciplines.

The MMLU benchmark consists of multiple-choice questions that require both broad knowledge and the ability to apply that knowledge to solve complex problems. It tests models not just on factual recall, but on their ability to reason, analyze, and draw conclusions across different fields of study. \\

\textbf{Tags:} \# Transparency

\subsection{Massive Multitask Language Understanding - Pro}
Massive Multitask Language Understanding-Pro (MMLU-Pro) \cite{abs-2406-01574} is a comprehensive benchmark enhancing and extending the mostly knowledge-driven MMLU \cite{HendrycksBBZMSS21} benchmark by integrating more challenging, reasoning-focused questions and expanding the choice set from 4 to 10 options. It spans 14 diverse domains including mathematics, physics, chemistry, law, engineering, psychology, and health, encompassing over 12,000 questions and thus meeting the breadth requirement for evaluation of multi-task language understanding capabilities in LLMs. \\

\textbf{Tags:} \# Transparency

\subsection{Google-Proof Q\&A Benchmark}
Google-Proof Q\&A Benchmark (GPQA) \cite{abs-2311-12022} is a challenging dataset 448 multiple-choice questions written by domain experts in biology, physics, and chemistry. The questions are high-quality and extremely difficult, with 74\% estimated objectivity on the basis of expert assessment, and 34\% accuracy by highly skilled, resourced, and motivated non-experts. \\

\textbf{Tags:} \# Transparency \# Diversity, non-discrimination and fairness

\subsection{Multistep Soft Reasoning}
Multistep Soft Reasoning (MuSR) \cite{SpragueYBCD24} is a reasoning dataset for evaluating LLMs on multistep soft reasoning tasks specified in a natural language (from either murder mysteries, object placement questions, or team allocation domains) using a novel neurosymbolic synthetic-to-natural generation algorithm,which can be scaled in complexity as more powerful models emerge. \\

\textbf{Tags:} \# Transparency

\subsection{Mathematics Aptitude Test of Heuristics}
Mathematics Aptitude Test of Heuristics (MATH) \cite{HendrycksBKABTS21} is a benchmark/dataset consisting of 12 500 problems from high school math competitions that measures the problem-solving ability of LLMs. Each problem has a step-by-step solution, final boxed answer, a difficulty tag from 1 to 5, and is one of the 7 subjects like Prealgebra, Algebra, Number Theory, Counting and Probability, Geometry, Intermediate Algebra, and Precalculus. \\

\textbf{Tags:} \# Technical Robustness and Safety
\# Transparency

\subsection{Instruction Following Evaluation}
Instruction Following Evaluation (IFEval) \cite{abs-2311-07911} is a benchmark for evaluating the proficiency of large language models in instruction following. It consists of a list of 25 verifiable instructions and a set of 541 prompts, with each prompt containing one or multiple verifiable instructions. Also 4 strict accuracy scores are defined to evaluate each model. \\

\textbf{Tags:} \# Technical Robustness and Safety
\# Transparency

\subsection{Big Bench Hard}
Big Bench Hard (BBH) \cite{SuzgunSSGTCCLCZ23} is a diverse evaluation suite of 23 challenging tasks (27 sub-tasks and in total 6511 evaluation examples) in different categories such as traditional NLP, mathematics, commonsense reasoning, and question-answering with the goal to test the capabilities of large language models using objective metrics. \\

\textbf{Tags:} \# Technical Robustness and Safety
\# Transparency

\subsection{Cord-19}
CORD-19 \cite{abs-2004-10706} is a large collection of publications and pre-prints on COVID-19 and related historical coronaviruses such as SARS and MERS. It now consists of over 140K papers with over 72K full texts, with papers in Medicine (55\%), Biology (31\%), and Chemistry (3\%), which together constitute almost 90\% of the corpus. CORD-19 was designed to facilitate the development of text mining and information retrieval systems over its rich collection of metadata and structured full text papers. \\

\textbf{Tags:} \# Dataset

\subsection{LAMBADA}
LAMBADA dataset \cite{PapernoKLPBPBBF16} is a dataset to evaluate the capabilities of computational models for text understanding by means of a word prediction task. It consists of 10,022 passages, divided into 4,869 development and 5,153 test passages with an average passage consisting of 4.6 sentences in the context plus 1 target sentence, for a total length of 75.4 tokens (dev)/ 75 tokens (test). Preliminary experiments suggest that even some cutting-edge neural network approaches that are in principle able to track long-distance effects are far from passing the LAMBADA challenge. \\

\textbf{Tags:} \# Technical Robustness and Safety
\# Transparency

\subsection{TriviaQA}
TriviaQA \cite{JoshiCWZ17} is a reading comprehension dataset consisting of high percentage of challenging questions with substantial syntactic and lexical variability and often requiring multi-sentence reasoning. TriviaQA contains over 650K question-answer-evidence triples, that are derived by combining 95K Trivia enthusiast authored question-answer pairs with on average six supporting evidence documents per question coming from two domains - Web search  results and Wikipedia pages. TriviaQA also provides a benchmark for a variety of other tasks such as IR-style question answering, QA over structured KBs and joint modeling of KBs and text, with much more data than previously available. \\

\textbf{Tags:} \# Transparency 

\subsection{WinoGrande}
WinoGrande \cite{SakaguchiBBC20} is a large-scale dataset of 44k problems, inspired by the original Winograd Schema Challenge (WSC) design (a set of 273 expert-crafted pronoun resolution problems originally designed to be unsolvable for statistical models that rely on selectional preferences or word associations), but adjusted to improve both the scale and the hardness of the dataset. The key steps of the dataset construction consist of (1) a carefully designed crowdsourcing procedure, followed by (2) systematic bias reduction using a novel AFLITE algorithm that generalizes human-detectable word associations to machine-detectable embedding associations. WINOGRANDE as a resource, demonstrates effective transfer learning and achieve state-of-the-art results on several related benchmarks. \newline

\textbf{Tags:} \# Technical Robustness and Safety
\# Transparency \# Diversity, non-discrimination and fairness

\subsection{CausalBench}
CausalBench \cite{wang-2024-causalbench} is a comprehensive benchmark for evaluating the causal reasoning capabilities of LLMs. It comprises four perspectives of causal reasoning for each scenario: cause-to-effect, effect-to-cause, cause-to-effect with intervention, and effect-to-cause with intervention. CausalBench includes a diverse set of problem types spanning textual, mathematical, and coding domains, enabling a comprehensive assessment of causal reasoning abilities across different modalities. The benchmark consists of more than 60,000 problems and employs six evaluation metrics to measure LLMs’ causal reasoning performance. \\

\textbf{Tags:} \# Technical Robustness and Safety
\# Transparency \# Diversity, non-discrimination and fairness

\subsection{CausalBench (2)}
CasulaBench(2) \cite{abs-2404-06349} is a comprehensive benchmark that encompasses three causal learning-related tasks: to identify correlation, causal skeleton, and causality. It incorporates 15 commonly used real-world causal learning datasets of diverse sizes, enabling comprehensive comparisons of LLMs’ performance with the classic causal learning algorithms. CausalBench uses four distinct prompt formats, which include one or more elements of variable names, background knowledge, and structured data and covers causal discovery datasets of various scales, ranging from 5 to 109 nodes, far exceeding what current evaluation works have explored. \\

\textbf{Tags:} \# Technical Robustness and Safety
\# Transparency \# Diversity, non-discrimination and fairness

\subsection{BackdoorLLM}
BackdoorLLM \cite{li2024backdoorllm} is a comprehensive benchmark for studying backdoor attacks on LLMs with a repository designed to facilitate research on backdoor attacks in LLMs. It includes a standardized pipeline for training backdoored LLMs with diverse strategies, such as data poisoning, weight poisoning, hidden state steering, and chain-of-thought attacks. The study (with over 200 experiments on 8 attacks across 7 scenarios and 6 model architectures) provides new insights into the nature of backdoor vulnerabilities in LLMs, which will aid in developing future defense methods against LLM backdoor attacks. \\

\textbf{Tags:} \# Technical Robustness and Safety
\# Transparency

\subsection{ConflictBank}
ConflictBank \cite{su2024conflictbankbenchmarkevaluatinginfluence} is a novel and comprehensive developed to systematically evaluate knowledge conflicts from three aspects: (i) conflicts encountered in retrieved knowledge, (ii) conflicts within the models’ encoded knowledge, and (iii) the interplay between these conflict forms. It contains a large diverse dataset of 553K QA pairs and 7M knowledge conflict evidence in high quality. The study also presents in-depth pilot experiments on twelve LLMs across four model series and provide comprehensive analyses about model scales, conflict causes, and conflict types. \\

\textbf{Tags:}  \# Diversity, non-discrimination and fairness

\subsection{Reefknot}
Reefknot \cite{zheng2024reefknotcomprehensivebenchmarkrelation} is a comprehensive benchmark called Reefknot to evaluate and mitigate relation hallucinations in multimodal large language models (MLLMs). Its dataset is constructed from over 20k data through a scene graph-based construction pipeline, covering two discriminative tasks (Y/N and MCQ) and one generative task (VQA). The paper also proposes a Detect-then-Calibrate method to mitigate the relation hallucination via entropy threshold, with an average reduction of 9.75\% in the hallucination rate across Reefknot and two other representative relation hallucination datasets. \\

\textbf{Tags:}  \# Technical Robustness and safety \# Diversity, non-discrimination and fairness

\subsection{DQA}
DQA \cite{zheng2024revolutionizingdatabaseqalarge} is the first comprehensive database Q\&A benchmark (DQA) consisting of a large-scale dataset with 240,000 Q\&A pairs. It also proposes a plug-and-play testbed encapsulating all components potentially involved in the database Q\&A, such as Question-Categorization Routing (QCR), Prompt-Template Engineering (PTE), Retriever-Augmented Generation (RAG) and Tool-Invocation Generation (TIG). Using DQA, the study also conducted a comprehensive evaluation to showcase DB Q\&A ability of seven general-purpose LLMs and two variants based on pre-training and fine-tuning evaluates. \\

\textbf{Tags:} \# Dataset \# Transparency

\subsection{MultiTrust}
MultiTrust \cite{abs-2406-07057} is a comprehensive benchmark designed to evaluate the trustworthiness of Multimodal Large Language Models (MLLMs). The benchmark covers five key aspects of trustworthiness: truthfulness, safety, robustness, fairness, and privacy. It employs a rigorous evaluation strategy that addresses both multimodal risks (new risks introduced by the visual modality) and cross-modal impacts (how the visual modality affects the performance on original text tasks). The authors conducted extensive experiments on 21 MLLMs, revealing that open-source models, despite their progress in general capabilities, still lag behind proprietary models in trustworthiness. The benchmark also highlights the detrimental effects of multimodality on MLLMs' trustworthiness, emphasizing the need for further research to enhance their reliability. To facilitate future research, the authors release a scalable toolbox for standardized trustworthiness evaluation. \newline

\textbf{Tags:} \# Technical Robustness and safety
\# Privacy and data governance
\# Transparency
\# Diversity, non-discrimination and fairness

\subsection{LTLBench}
LTLBench \cite{abs-2407-05434} is a benchmark designed to evaluate the temporal reasoning capabilities of Large Language Models (LLMs). Temporal reasoning, crucial for AI understanding of event sequences and relationships, is assessed in LLMs using various datasets. The authors propose a novel pipeline for constructing such datasets, leveraging random directed graphs, Linear Temporal Logic (LTL) formulas, and the NuSMV model checker. This approach allows for the generation of diverse and scalable temporal reasoning problems. The authors used this pipeline to create LTLBench, a dataset of 2,000 temporal reasoning challenges, and evaluated six LLMs on it. Results show LLMs exhibit promise in handling temporal reasoning but still struggle with complex scenarios. The work contributes a valuable tool for evaluating and improving the temporal reasoning abilities of LLMs and other AI systems.\newline

\textbf{Tags:} \# Technical Robustness and Safety
\# Transparency
\# Accountability

\textbf{Kewords}: \textit{Large Language Models, Temporal Logic, Reasoning }

\subsection{Large Langugage Model in the Clinic}
ClinicBench \cite{liu2024largelanguagemodelsclinic} is a comprehensive benchmark designed to evaluate large language models (LLMs) in clinical settings. The benchmark includes a diverse set of tasks, including both traditional machine learning tasks and novel, clinically relevant tasks such as referral question answering, treatment recommendation, and patient education generation. The authors evaluate a range of LLMs on these tasks, finding that while models like GPT-4 show promising performance on certain tasks, there remains a significant gap between current LLM capabilities and the requirements for real-world clinical application. The paper also explores the impact of different types of fine-tuning data on model performance, highlighting the potential benefits of incorporating clinical-standard knowledge bases into the training process.\newline

\textbf{Tags:} \# Technical Robustness and Safety
\# Transparency

\subsection{High-Quality Hallucination Benchmark}
The High-Quality Hallucination (HQH) \cite{abs-2406-17115} Benchmark addresses the problem of hallucination in Large Vision-Language Models (LVLMs), where models generate text that's inconsistent with the visual input. The authors point out that existing benchmarks for evaluating this issue suffer from varying quality, impacting the reliability of model assessments. To tackle this, they propose a framework called HQM to measure the quality of these benchmarks based on reliability and validity. Their analysis reveals that current benchmarks have limitations in both aspects. \newline

In response, they develop a new benchmark, HQH, which uses free-form Visual Question Answering and a simplified evaluation metric to enhance its quality. Their evaluation on various LVLMs demonstrates that hallucination remains a challenge, particularly with certain types like existence, OCR, and comparison. The paper concludes by emphasizing the need for further research to mitigate hallucination in LVLMs and advocating for the use of their HQM framework to improve the quality of future benchmarks.\newline

\textbf{Tags:} \# Technical Robustness and Safety
\# Transparency
\# Accountability

\subsection{AI Benchmarking for Science}
The MLCommons Science Working Group \cite{ThiyagalingamLY22} is creating science-specific AI benchmarks to advance AI's application in scientific research. They've developed four benchmarks so far, each with datasets and reference implementations, covering areas like atmospheric science, material science, healthcare, and earth science.

\begin{itemize}
    \item \textbf{Cloud Masking} (cloud-mask): Classifies pixels in satellite images as containing cloud or clear sky, crucial for estimating sea surface temperature.
    \item \textbf{Space Group Classification of Solid State Materials} (stemdl): Classifies the space group of solid-state materials from electron diffraction patterns, aiding in understanding material properties.
    \item \textbf{Time Evolution Operator} (tevelop): Predicts the evolution of time series data, exemplified by earthquake forecasting.
    \item \textbf{Predicting Tumor Response to Single and Paired Drugs} (candle-uno): Predicts tumor response to drug treatments based on molecular features.
\end{itemize}

These benchmarks help evaluate AI/ML techniques for scientific problems, considering factors like performance, explainability, and scalability. The group emphasizes clear benchmarking policies and plans to expand its benchmark collection, ensuring datasets are FAIR compliant and encouraging community contributions. The goal is to facilitate AI adoption in science, aiding algorithm selection, fostering collaboration, and accelerating discoveries.\newline

\textbf{Tags:} \# Technical Robustness and Safety
\# Transparency
\# Societal and Environmental Well-being
\# Accountability

\subsection{TPCx-AI}
TPCx-AI \cite{BruckeHPPR23} is an industry-standard benchmark designed to evaluate the end-to-end performance of artificial intelligence (AI) and machine learning (ML) systems. It emphasizes that TPCx-AI is unique in its focus on evaluating the entire ML pipeline, from data ingestion to model serving and post-processing, unlike other benchmarks that primarily focus on model training. TPCx-AI is built on a retail scenario and includes ten diverse use cases, covering both traditional ML and deep learning models. It provides a comprehensive toolkit with implementations in Python and Apache Spark, making it readily usable for evaluating ML/AI systems on standard hardware. The benchmark also includes a scalable dataset and a well-defined performance metric, ensuring fair and meaningful comparisons between different systems. \newline

\textbf{Tags:} \# Technical Robustness and Safety
\# Transparency
\# Accountability

\subsection{RobustBench}
RobustBench \cite{CroceASDFCM021} is a standardized benchmark for evaluating the adversarial robustness of machine learning models. It emphasizes the need for accurate and reliable robustness evaluations, addressing the challenge of robustness overestimation that often hinders progress in the field. RobustBench focuses on image classification tasks and establishes restrictions on the allowed models to ensure meaningful comparisons. It utilizes AutoAttack, an ensemble of white- and black-box attacks, as the standard evaluation method. The benchmark also encourages external evaluations using adaptive attacks to further enhance its reliability. RobustBench provides a leaderboard showcasing the performance of various models on well-defined tasks in different threat models and on common corruptions. Additionally, it offers a Model Zoo, a collection of robust models readily available for downstream applications. Overall, RobustBench aims to accelerate progress in adversarial robustness research by providing a standardized and reliable benchmark, facilitating the identification of the most promising ideas in training robust models.\newline

\textbf{Tags:} \#Technical Robustness and Safety
\#Accountability
\#Transparency

\subsection{MetaBox}
MetaBox \cite{MaGCLPG0C23} is the first benchmark platform specifically designed for Meta-Black-Box Optimization with Reinforcement Learning (MetaBBO-RL). MetaBBO-RL aims to automate the fine-tuning of black-box optimizers, thereby enhancing their performance across various problem instances. The platform offers a flexible algorithmic template, a diverse collection of over 300 problem instances, and an extensive library of 19 baseline methods. It also introduces three standardized performance metrics for a more comprehensive evaluation of MetaBBO-RL methods. MetaBox is open-source and accessible on GitHub, contributing to the advancement of research in this field.\newline

\textbf{Tags:} \#Technical Robustness and Safety
\#Accountability
\#Transparency

\subsection{SUC}
Structural Understanding Capabilities (SUC) \cite{sui2024tablemeetsllmlarge} is a comprehensive benchmark consisting of seven tasks, each with its own unique challenges, e.g., cell lookup, row retrieval, and size detection. The authors try to answer the following questions: 1) What input designs and choices are most effective in enabling LLMs to understand tables?; 2) To what extent do LLMs already possess structural understanding capabilities for structured data? The comparison reveals that LLMs have the basic capabilities towards understanding structural information of tables and to further improve it propose self-augmentation for effective structural prompting, such as critical value / range identification using internal knowledge of LLMs.\newline

\textbf{Tags:} \# Technical Robustness and Safety
\# Transparency

\textbf{Kewords}: \textit{Large Language Models, Semi-structured Data, Structural Understanding Capabilities}

\subsection{LogicVista}
LogicVista \cite{abs-2407-04973} is a benchmark designed to assess the logical reasoning skills of Multimodal Large Language Models (MLLMs). It focuses on evaluating how well these models can reason and solve problems based on visual information. The benchmark uses a variety of visual inputs, including diagrams, text, patterns, graphs, tables, 3D shapes, puzzles, and sequences. LogicVista tests five types of logical reasoning: deductive, inductive, numerical, spatial, and mechanical.  It provides a tool for evaluating how well these models can apply logical reasoning skills across different types of visual data. The project is open source and available on GitHub.\newline

\textbf{Tags:} \# Technical Robustness and Safety

\subsection{RelevAI-Reviewer}
RelevAI-Reviewer \cite{abs-2406-10294} is a new AI tool for automatically reviewing scientific papers. The tool is designed to overcome the limitations of traditional peer review processes, which can be slow, biased, and inconsistent. RelevAI-Reviewer specifically focuses on evaluating the relevance of survey papers based on a given prompt, similar to a "call for papers."

The authors created a dataset of over 25,000 survey papers from 22 different fields, categorized into four relevance levels. They tested various machine learning models, including Support Vector Machines (SVM) and BERT, to determine the most effective approach. The results showed that BERT models, particularly with a specific encoding method called "thermometer encoding," outperformed other methods in accurately ranking papers by relevance.\newline

\textbf{Tags:} \# Technical Robustness and Safety
\# Accountability
\# Transparency

\subsection{AI Safety Benchmark v0.5}
The AI Safety Benchmark v0.5 \cite{abs-2404-12241} is a proof-of-concept for evaluating the safety risks of AI systems, specifically chat-tuned language models. It presents a taxonomy of 13 hazard categories and includes tests for seven of them, using over 43,000 test items (prompts) to assess model responses. The benchmark aims to provide a standardized and interpretable evaluation of AI safety, addressing a critical need as AI systems become increasingly integrated into various domains. The authors acknowledge limitations in the current version and welcome feedback for the development of the full v1.0 benchmark planned for the end of 2024.\newline

\textbf{Tags:} \# Technical Robustness and Safety
\# Accountability 
\# Diversity, non-discrimination and fairness

\subsection{Dataset OpenAssistant Conversations}
OpenAssistant Conversations \cite{abs-2304-07327} is a large collection of human-generated and human-annotated assistant-style conversations. This dataset was created by over 13,500 volunteers with the goal of democratizing research on large-scale language model alignment. The dataset consists of 161,443 messages in 35 different languages, annotated with 461,292 quality ratings, resulting in over 10,000 complete conversation trees. The authors used this dataset to fine-tune several language models and found that they show consistent improvements on standard benchmarks over their respective base models. The dataset is released under a fully permissive license and is available on the Hugging Face Hub.\newline

\textbf{Tags:} \# Dataset

\subsection{ZebraLogic}
ZebraLogic Benchmark \cite{ZebraLogic} is based on Logic Grid Puzzles, to test how well LLMs can reason. The puzzles involve houses with distinct features that need to be deduced from clues. The article also explores different metrics for evaluating the LLMs' performance, including puzzle-level accuracy and cell-wise accuracy. It was found that LLMs struggle with complex logical reasoning tasks, particularly those that require counterfactual thinking, reflective reasoning, and structured memorization. The article also provides an example of a 2x3 puzzle and explains the evaluation method. Some of the important points are that LLMs are still weak in logical reasoning tasks.\newline

\textbf{Tags:} \# Technical Robustness and Safety
\# Transparency 

\subsection{GLoRE Benchmark}
GLoRE \cite{abs-2310-09107} is a benchmark for evaluating the logical reasoning abilities of LLMs, comprising 12 datasets over three types of tasks, including Multi-choice Reading Comprehension, Natural Language Inference (NLI), and True-or-False (TF) questions. Using GLoRE, the authors evaluate the logical reasoning abilities of several LLMs, including ChatGPT, GPT-4, and open-source LLMs based on LLaMA and Falcon. The results indicate that both ChatGPT and GPT-4 outperform open-source LLMs and traditional supervised fine-tuned models in most tasks. However, the performance of these models is not consistent across datasets, indicating their sensitivity to data distribution. The paper also proposes a self-consistency probing method to enhance the accuracy of ChatGPT and a fine-tuned method to boost the performance of an open LLM.\newline

\textbf{Tags:} \# Technical Robustness and Safety

\textbf{Kewords}: \textit{Large Language Model, Logical Reasoning}

\subsection{General Language Understanding Evaluation}
The General Language Understanding Evaluation (GLUE) \cite{wang2019gluemultitask} benchmark provides a comprehensive platform for evaluating and analyzing natural language understanding (NLU) systems. Designed to foster the development of generalizable and robust NLU models, GLUE offers a suite of nine tasks, including single-sentence classification, similarity and paraphrase detection, and natural language inference. The tasks were selected to cover a range of genres, dataset sizes, and challenges. GLUE also features a diagnostic dataset for analyzing specific linguistic capabilities of models, including lexical semantics, logical reasoning, and predicate-argument structure.

The GLUE benchmark is model-agnostic, allowing the use of any architecture capable of processing single sentences or sentence pairs. Its primary evaluation relies on performance across all tasks, encouraging the development of models that effectively transfer knowledge and perform well even with limited training data. To ensure fairness and prevent overfitting, test labels for some tasks are privately held and results must be submitted to an online platform for evaluation. \\

\textbf{Tags:} \# Technical Robustness and Safety
\# Transparency
\# Diversity, non-discrimination and fairness

\subsection{A Stickier Benchmark for General-Purpose Language Understanding Systems}
SuperGLUE (A Stickier Benchmark for General-Purpose Language Understanding Systems) \cite{wang2020superglue} is a benchmark designed to address the limitations of its predecessor, GLUE \cite{wang2019gluemultitask}, by introducing more challenging natural language understanding (NLU) tasks. The motivation for SuperGLUE stems from significant advances in NLU models, such as BERT and OpenAI GPT, which have largely surpassed human performance on GLUE tasks. This progress highlighted the need for a more robust and challenging evaluation framework to continue driving innovation.

Key Features of SuperGLUE:
\begin{itemize}
    \item Task Design: SuperGLUE consists of eight tasks that emphasize reasoning, understanding, and contextual comprehension. These tasks include:
     \begin{itemize}
         \item BoolQ: Yes/no question answering from text.
         \item CB: CommitmentBank for entailment in embedded clauses.
         \item COPA: Choice of plausible alternatives requiring causal reasoning.
         \item MultiRC: Multi-sentence reading comprehension with multiple correct answers.
         \item ReCoRD: Cloze-style questions demanding commonsense reasoning.
         \item RTE: Recognizing textual entailment.
        \item WiC: Word-in-context sense disambiguation.
        \item WSC: The Winograd Schema Challenge for coreference resolution.
     \end{itemize}
    \item Diversity in Task Formats: Unlike GLUE, which primarily involved sentence and sentence-pair classification, SuperGLUE incorporates formats like coreference resolution and multi-answer question answering to challenge existing models.
    \item Human Baselines: SuperGLUE provides comprehensive human performance baselines for all tasks, ensuring clear headroom for machine models to improve.
    \item Diagnostic Tools: Includes a diagnostic dataset to assess linguistic, commonsense, and world knowledge.
    \item Software and Leaderboard: Offers a modular toolkit built around PyTorch and AllenNLP for easy use and an online leaderboard for tracking progress.
    \item Scoring and Evaluation: Combines performance on all tasks into a single score, balancing metrics to provide a holistic view of system capabilities.
\end{itemize}

\textbf{Tags:} \# Technical Robustness and Safety
\# Transparency
\# Diversity, non-discrimination and fairness

\subsection{AI2 Reasoning Challenge}
The AI2 Reasoning Challenge (ARC) \cite{Yadav_2019} dataset is a benchmark designed to assess and advance the capabilities of question-answering (QA) systems in handling scientific reasoning. It includes multiple-choice questions sourced from U.S. grade-school science exams, split into two partitions:
\begin{itemize}
    \item \textbf{Easy Set}: Contains questions that can be answered by simple retrieval or basic reasoning.
    \item \textbf{Challenge Set}: Features more complex questions that require reasoning across multiple sentences or external knowledge sources.
\end{itemize}

Key Features:
\begin{itemize}
    \item \textbf{Complexity}: The Challenge Set requires multi-hop reasoning, commonsense knowledge, and understanding of scientific principles, posing a significant difficulty for existing QA models.
    \item \textbf{Scale}: The ARC dataset includes thousands of questions, with four answer choices per question, covering grades 3 to 9 science topics.
    \item \textbf{Knowledge Base}: Accompanying the dataset is a corpus of 14.3 million unstructured text passages, intended as a resource for models to retrieve evidence for answering questions.
    \item \textbf{Evaluation Metrics}: Standard accuracy metrics are used to evaluate QA performance.
\end{itemize}

\textbf{Tags:} \# Transparency

\subsection{Natural Language for Visual Reasoning for Real}
The Natural Language for Visual Reasoning for Real (NLVR2) \cite{suhr2019corpus} benchmark introduces a dataset designed for joint reasoning about natural language and images, focusing on semantic diversity, compositionality, and visual reasoning challenges. Comprising 107,292 examples of English sentences paired with web photographs, the task requires determining whether a caption is true for a given pair of images.

This dataset was crowd-sourced using visually rich images and a compare-and-contrast task, ensuring diverse linguistic output. Unlike previous resources that often relied on synthetic data, NLVR2 emphasizes real-world images, which enhances the complexity and richness of the language used.

NLVR2 demonstrates significant challenges for state-of-the-art visual reasoning methods, as evaluations showed relatively low performance on this benchmark, indicating its robustness. The dataset includes 29,680 unique sentences and 127,502 images, and is structured to facilitate various reasoning tasks, making it a valuable resource for advancing research in natural language understanding and visual reasoning. \\

\textbf{Tags:} \# Transparency

\subsection{Visual Question Answering }
The Visual Question Answering (VQA) \cite{marino2019okvqa} is benchmark that emphasizes knowledge-based reasoning beyond visual content. OK-VQA comprises over 14,000 questions that necessitate external knowledge for accurate answers, thereby encouraging the development of models capable of reasoning and retrieving information from outside sources.

The dataset includes a diverse range of questions across various knowledge categories such as history, science, and geography. The authors detail their methodology for dataset creation, which involves filtering low-quality questions and ensuring a uniform distribution of answers to mitigate biases present in previous datasets. OK-VQA is designed to challenge existing VQA systems by requiring them to integrate visual understanding with external knowledge retrieval, thus promoting advancements in the field. \\

\textbf{Tags:} \# Technical Robustness and Safety
\# Transparency

\section{Summary}

Table \ref{tab:benchmark_datasets} presents a comprehensive list of AI benchmarks and datasets, along with the seven EU requirements for Trustworthy AI that each addresses: Human Agency and Oversight [\textbf{HAO}], Technical Robustness and Safety [\textbf{TRS}], Privacy and Data Governance [\textbf{PDG}], Transparency [\textbf{T}],  Diversity, Non-discrimination and Fairness [\textbf{DNF}], Societal and Environmental Well-being [\textbf{SEW}] and Accountability [\textbf{A}].

\begin{table}[htbp]
\centering
\caption{Overview of Benchmarks and Datasets}
\label{tab:benchmark_datasets}
\begin{tabular}{|l|c|c|c|c|c|c|c|}
\hline
\textbf{Benchmark / Dataset} & \textbf{HAO} & \textbf{TRS} & \textbf{PDG} & \textbf{T} & \textbf{DNF} & \textbf{SEW} & \textbf{A} \\ \hline
ANLI \cite{NieWDBWK20} & & y & & & & & \\ \hline
HellaSwag \cite{ZellersHBFC19} & & & &  y & & & \\ \hline
CommonsenseQA \cite{TalmorHLB19} & & & & & & & \\ \hline
MMLU \cite{HendrycksBBZMSS21} & & & & y & & & \\ \hline
MMLU-Pro \cite{abs-2406-01574} & & & & y & & & \\ \hline
GPQA \cite{abs-2311-12022} & &  &  & y & y & & \\ \hline
MuSR \cite{SpragueYBCD24} & &  & & y & & & \\ \hline
MATH \cite{HendrycksBKABTS21} & & y & & y & & &  \\ \hline
IFEval \cite{abs-2311-07911} & & y & & y & & &  \\ \hline
BBH \cite{SuzgunSSGTCCLCZ23} & & y & & y & & &  \\ \hline
CORD-19 \cite{abs-2004-10706} & & & & & & & \\ \hline
LAMBADA \cite{PapernoKLPBPBBF16} & & y & & y & & &  \\ \hline
TriviaQA \cite{JoshiCWZ17} & & & & y & & &  \\ \hline
WinoGrande \cite{SakaguchiBBC20} & & y & & y & y & &  \\ \hline
CausalBench \cite{wang-2024-causalbench} & &y & & y & y & &  \\ \hline
CasulaBench(2) \cite{abs-2404-06349} & & y & & y & y & &  \\ \hline
BackdoorLLM \cite{li2024backdoorllm} & & y & & y & & &  \\ \hline
ConflictBank \cite{su2024conflictbankbenchmarkevaluatinginfluence} & & & & &  y & & \\ \hline
Reefknot \cite{zheng2024reefknotcomprehensivebenchmarkrelation} &  & y & & & y & & \\ \hline
DQA \cite{zheng2024revolutionizingdatabaseqalarge} & &  & & y & & & \\ \hline
MultiTrust \cite{abs-2406-07057} & & y & y & y & y & &  \\ \hline
LTLBench \cite{abs-2407-05434} & & y & & y & & & y  \\ \hline
ClinicBench \cite{liu2024largelanguagemodelsclinic} & & y & & y & & &  \\ \hline
HQH \cite{abs-2406-17115} & & y & & y & & & y  \\ \hline
MLCommons Science WG \cite{ThiyagalingamLY22} & & y & & y & & y & y \\ \hline
TPCx-AI \cite{BruckeHPPR23} & & y & & y & & &  y\\ \hline
RobustBench \cite{CroceASDFCM021} & & y & & y & & & y  \\ \hline
MetaBox \cite{MaGCLPG0C23} & &  y & & y & &  & y \\ \hline
SUC \cite{sui2024tablemeetsllmlarge} & & y & & y & &  &  \\ \hline
LogicVista \cite{abs-2407-04973} & & y & & & & & \\ \hline
RelevAI-Reviewer \cite{abs-2406-10294} & & y & & y & & & y  \\ \hline
AI Safety Benchmark v0.5 \cite{abs-2404-12241} & & y & & y & y& &   \\ \hline
OpenAssistant Conversations \cite{abs-2304-07327} & & & & & & & \\\hline
ZebraLogic \cite{ZebraLogic} & & y & & y & & &  \\ \hline
GLoRE \cite{abs-2310-09107} & & y & & & & &  \\ \hline
GLUE \cite{wang2019gluemultitask} & & y & & y & y &  &  \\ \hline
SuperGLUE \cite{wang2020superglue} & & y & & y &  y & &  \\ \hline
ARC \cite{Yadav_2019} & & & &  y &  & & \\ \hline
NLVR2 \cite{suhr2019corpus} & & & & y &  & & \\ \hline
VQA \cite{marino2019okvqa} & & y & & y & & &  \\ \hline
\end{tabular}
\end{table}

\newpage

\bibliographystyle{plain}
\bibliography{bibliography.bib}

\end{document}